# A Topological Representation for Taking Cities as a Coherent Whole

Bin Jiang

Faculty of Engineering and Sustainable Development, Division of GIScience
University of Gävle, SE-801 76 Gävle, Sweden
Email: bin.jiang@hig.se

*(Draft: May 2016, Revision: July 2016, January, May, August 2017)*

*"It is possible that one day computer programs designed for cognition might be able to pick out these centers and rank order them by their degrees of life."*

Alexander (2002-2005, Book 1, p. 365)

**Abstract**
A city is a whole, as are all cities in a country. Within a whole, individual cities possess different degrees of wholeness, defined by Christopher Alexander as a life-giving order or simply a living structure. To characterize the wholeness and in particular to advocate for wholeness as an effective design principle, this paper develops a geographic representation that views cities as a whole. This geographic representation is topology-oriented, so fundamentally differs from existing geometry-based geographic representations. With the topological representation, all cities are abstracted as individual points and put into different hierarchical levels, according to their sizes and based on head/tail breaks - a classification and visualization tool for data with a heavy tailed distribution. These points of different hierarchical levels are respectively used to create Thiessen polygons. Based on polygon-polygon relationships, we set up a complex network. In this network, small polygons point to adjacent large polygons at the same hierarchical level and contained polygons point to containing polygons across two consecutive hierarchical levels. We computed the degrees of wholeness for individual cities, and subsequently found that the degrees of wholeness possess both properties of differentiation and adaptation. To demonstrate, we developed four case studies of all China and UK natural cities, as well as Beijing and London natural cities, using massive amounts of street nodes and Tweet locations. The topological representation and the kind of topological analysis in general can be applied to any design or pattern, such as carpets, Baroque architecture and artifacts, and fractals in order to assess their beauty, echoing the introductory quote from Christopher Alexander.

## 1. Introduction

Geographic representation or representing the Earth's surface constitutes one of the main research areas in geographic information systems (GIS) and science. Existing geographic representations, such as raster and vector, are essentially geometry-based because of involved locations, directions, and sizes. These representations are mainly driven by a static map metaphor (Yuan et al. 2005, Goodchild, Yuan and Cova 2007) for representing and characterizing the Earth's surface. They can be used to measure things and to assess the relationship of things at local scales, such as spatial dependence. However, things are not measurable, or measurement depends on the measuring scales because of the fractal nature of geographic phenomena (e.g., Goodchild and Mark 1984, Batty and Longley 1994, Chen 2011). Spatial dependence is just one of two spatial properties. The other is spatial heterogeneity across all scales. Geometry-based representations fail to capture the spatial property of heterogeneity across all scales (Jiang, Zhao and Yin 2008). To better understand geographic forms or cities' structure, we must rely on topology-based representations, e.g., the topological representation of named or natural streets (Jiang and Claramunt 2004, Jiang, Zhao and Yin 2008). This paper develops another topological representation that views cities as a whole. We refer to topology as an instrument that enables us to see the underlying spatial heterogeneity, or scaling property of far more small things than large ones. The topology used in GIS, however, is essentially geometry-based, referring to adjacent relationships of basic geometric elements such as points, lines, polygons and pixels.

The geometry-based topology is a key principle for defining efficient data structure with adjacent relationships stored in databases, which can significantly reduce data storage. The topology or planar topology is a rigorous method for identifying digitizing errors such as self-intersecting polygons, islands, overshoots and undershoots, and slivers (Corbett 1979, Theobald 2001). One typical example is the topologically integrated geographic encoding and referencing (TIGER), which was developed by the US Census Bureau in the 1970s. With the TIGER, adjacent geometric elements, such as points, lines and polygons, are clearly defined and stored in the data structure. For example, two polygons can have one of eight possible relationships: Disjoint, contains, inside, equal, meet, covers, covered by, and overlap (Egenhofer and Herring 1990). Another use of topology in the GIS literature is with cartograms, in which geometric aspects such as distances and areas are dramatically distorted, but the underlying topological relationships are retained. The most well-known example of cartograms is the London Underground map, devised in 1933 by Harry Beck. Both topological data structures and cartograms still retain some geometry. In the London Underground map, the geographic locations of stations are still relatively correct, although the distances between stations are distorted. With both topological data structures and cartograms, one, however, cannot see the underlying scaling of far more small things than large ones; see Jiang, Zhao and Yin (2008) for details. This makes our use of topology unique, because the topology enables us to see the underlying scaling property or hierarchy.

We put forward a topological representation that takes cities as a coherent whole and enables us to see their underlying scaling hierarchy. This representation makes it possible for a configurational analysis based on the concept of wholeness (Alexander 2002–2005, Jiang 2015b). See Section 2 for an introduction to wholeness and its 15 fundamental properties. Based on the configurational analysis, we can see clearly that individual cities are differentiated from and adapt to each other to form a coherent whole. With the topological representation, we can compute the degrees of wholeness or beauty of individual cities for better understanding cities' structure. We develop some useful visualization methods (c.f., Section 4 for case studies) to illustrate the underlying scaling of cities in terms of their sizes and their degrees of wholeness. We further demonstrate the properties of differentiation and adaptation of cities, and provide eventually analytical evidence for and insights into the structure-preserving or wholeness-extending transformations (Alexander et al. 1987, Alexander 2002–2005) as the fundamental urban design principles.

Section 2 introduces the concept of wholeness or living structure, and its 15 geometric properties. Using 100 city sizes that strictly follow Zipf's law (1949), Section 3 illustrates the topological representation for viewing cities as a whole. To further demonstrate how the topological representation adds new insights into the cities' structure, Section 4 reports two major properties of differentiation and adaptation from case studies by applying the representation to China and UK natural cities. Section 5 further discusses the implications of the topological representation and wholeness for planning cities with a living structure. Finally Section 6 draws a conclusion and points out to future work.

**2. Wholeness or living structure, and its 15 fundamental properties**
Wholeness is defined as the structure that exists in space by all various coherent entities (called centers) and how these centers are nested in, and overlap with, each other (Alexander 2002–2005). Thus wholeness is a phenomenon that is inherent to space or matter, rather than just what is perceived as the gestalt (Köhler 1947). For the sake of simplicity, we use the Sierpinski carpet as a working example for introducing the concept of wholeness. The carpet is created iteratively, or generated, by removing one-ninth of the square of size 1 by 1 (Figure 1a). In theory, this process continues indefinitely, but we limited it here to three scales: 1/3, 1/9, and 1/27. The different squares of the carpet are not fragmented, but are a whole with a high degree of wholeness, called *living structure*. However, the carpet is not a typical living structure often seen in reality, given its strict definitions or exactness. It is strictly defined both globally and locally in terms of the square shapes. It is strictly defined by the constant, exact 1/3 scaling ratio. This strictness makes the carpet different from most living structures in reality, such as trees, mountains, coastlines, and cities. These real-world living structures are transformed – so called wholeness-extending transformations – under the guidance of the 15 properties, which are also called transformations (Table 1), whereas the carpet is created by following some strict mathematic rules. Despite of the strictness, the carpet shares the same spirit of wholeness or living structure. A living structure is a geometrical coherence made of, or more precisely transformed by, the 15 fundamental properties. As a living structure, the carpet possesses many of these properties, although it also misses some. The carpet contains one, eight, and 64 squares respectively at three levels of scale 1/3, 1/9, and 1/27, indicating that there are far more small squares than large ones. These squares are not fragmented but are a whole (Figure 1b).

Table 1: The 15 properties of the living structure or wholeness (Alexander 2002-2005)

| Levels of scale | Good shape | Roughness |
|---|---|---|
| Strong centers | Local symmetries | Echoes |
| Thick boundaries | Deep interlock and ambiguity | The void |
| Alternating repetition | Contrast | Simplicity and inner calm |
| Positive space | Gradients | Not separateness |

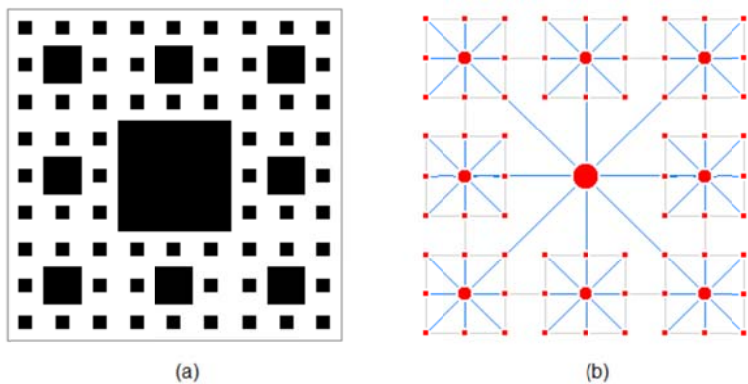

Figure 1: (Color online) The Sierpinski carpet acting or perceived as a whole
(Note: The Sierpinksi carpet with the three hierarchical scales: 1/3, 1/9 and 1/27 (a) is represented as a complex network consisting of many individual nodes as a whole (b), in which the dot sizes represent the degrees of wholeness, and the gray and blue lines indicate relationship respectively within a same scale and across two consecutive scales.)

The notion of far more small things than large ones refers to a recurring pattern so should be rephrased more truly as numerous smallest things, a few largest things, and some in between the smallest and the largest (Jiang and Yin 2014). The different scales form a continuum in the carpet and make it whole (Figure 1b), so it creates life. There are many strong centers in the carpet, creating a sensation of centeredness in a recursive manner. For example, the largest square is a strong center surrounded by the eight middle-sized squares, which constitute strong centers surrounded by the sixty-four smallest squares. Each square is surrounded by a boundary, but it is less obvious in the carpet. The property of alternating repetition is missing in the mathematical carpet, but it often appears in real carpets; see Alexander (1993) and Alexander (2002–2005) for more examples. In the carpet, the spaces between the black squares (the white part) are positive spaces, but are not as well-shaped as the black squares. In many good designs, such as the Nolli map of Rome, the ground space (the spaces between figural patterns) is as well-shaped as the figure space, forming positive spaces. The carpet is a good shape, since it consists of many squared shapes. The carpet is locally and globally symmetrical. This is true for most of real carpets. In the carpet, the local symmetries are all the same and uniform. This appearance lacks alternating repetition. However, a living structure is not necessarily globally symmetrical, but local symmetries are essential. The Alhambra plan, well studied by Alexander (2002-2005), is a typical example of local symmetries.

The property of deep interlock and ambiguity is clearly missing in the carpet, but it appears in many real carpets. The black and white contrast is obvious in the carpet. The property of gradient is less obvious with the carpet. Roughness is completely missing in the carpet because of the exactness of the shapes and scaling ratio. The property of echoes, in which the parts echo the whole, is shown in the carpet. The largest square is the void. Each of the eight middle-sized squares is surrounded by eight smallest squares, forming a kind of simplicity and inner calm, just as the eight middle-sized squares. None of the squares are separated from each other, but form a scaling hierarchy, which underlies the complex network as a whole (Figure 1b). With the complex network, there are two kinds of links: those among the same scales that are undirected or mutually directed; and those across two consecutive scales that are directed from smaller to larger ones. The property of not-separateness, or the 15 properties in general, advocate for a new worldview in which we see things in their wholeness, which underlies the topological representation we put forward in this paper. This new worldview differs fundamentally from the $20^{th}$-century mechanistic worldview. Wholeness evokes a sense of beauty that we can experience, life comes from wholeness, and eventually order emerges from wholeness. Wholeness is therefore the source of life, or beauty or order (or good design), which all can be interchangeably used. The wholeness that produces life and evokes beauty in buildings and cities is *"a direct result of the physical and mathematical structure that occurs in space, something which is clear and definite, and something which can be described and understood"* (Alexander 2002–2005, Book 1, p. 62). The physical and mathematical structure can be described as a hierarchical graph (Jiang 2015b), which helps address not only why a design is beautiful, but also how much beauty the design has. Our proposed topological representation aims to characterize the structure for cities as a whole.

**3. The topological representation for taking things as a whole**
The topological representation takes cities as a whole. To illustrate the representation, we create 100 cities that exactly follow Zipf's law (1949): 1, 1/2, 1/3, ... , and 1/100. The 100 cities are given some random locations in a two-dimensional space (Figure 2a). According to their sizes, the cities are put into different hierarchical levels based on head/tail breaks - a classification scheme and visualization tool for data with a heavy-tailed distribution (Jiang 2013, Jiang 2015a). The locations of the cities at the different hierarchical levels are respectively used to create Thiessen polygons (Figure 2b). Based on the polygon-polygon relationships, we set up a complex network, in which small polygons point to large ones at a same level, and contained polygons point to containing polygons across two consecutive levels (Figure 2c). With this complex network, we can compute the degrees of wholeness for individual cities (Jiang 2015b), which are represented by the dot sizes (Figure 2c). Unlike the city sizes, the degrees of wholeness are well adapted to their surroundings. It is clear that the cities statistically demonstrate the scaling property of far more small cities than large ones, since they follow Zipf's law (1949). However, they are not correctly arranged geometrically or according to central place theory (Christaller 1933, 1966), since the cities are given some random locations. Nevertheless, given this spatial configuration and according to the theory of centers (Alexander 2002-2005), individual cities obtain strength or wholeness from others as a whole. The computed degrees of wholeness or beauty are still put at the three hierarchical levels (Figure 2c).

There are essentially two types of beauty or harmony shown in the pattern of Figure 2c: One type among nearby things that are well adapted to each other with more or less similar sizes; and the other type across all scales that are differentiated from each other with far more small sizes than large ones. Comparing the initial city sizes and degrees of wholeness, we notice that some cities move up or down in the hierarchical levels, while some remain unchanged. Observing carefully, those cities that remain unchanged or move up have good support from others as a whole, while those that move down have less support. From this observation, we can understand that the degrees of wholeness are essentially exogenous, rather than endogenous. In other words, cities move up or down hierarchal levels because of the other cities in the interconnected whole, rather than just themselves. In fact, these cities moving up or down in the hierarchical levels can be directly linked to cities expanding and shrinking, which often occurs in reality. For example, the upper-most green city was growing, whereas the lower-most red city was shrinking. Cities' expansion and shrinkage, or urban structure and dynamics in general, can be better understood from the point of view of wholeness. Cities are not isolated, and support (and are supported by) others within a whole. In this regard, the degrees of wholeness can be thought of as ideal 'city sizes' that ought to be, under the spatial configuration. These city sizes are harmonized with surroundings, with differentiation across all scales and adaptation among those nearby. In the following case studies, we examine the city sizes and degrees of wholeness in three different ways: (1) Power law detection for these two metrics (differentiation), (2) calculation and visualization of ht-index (differentiation and adaptation), and (3) correlation between these two metrics (adaptation).

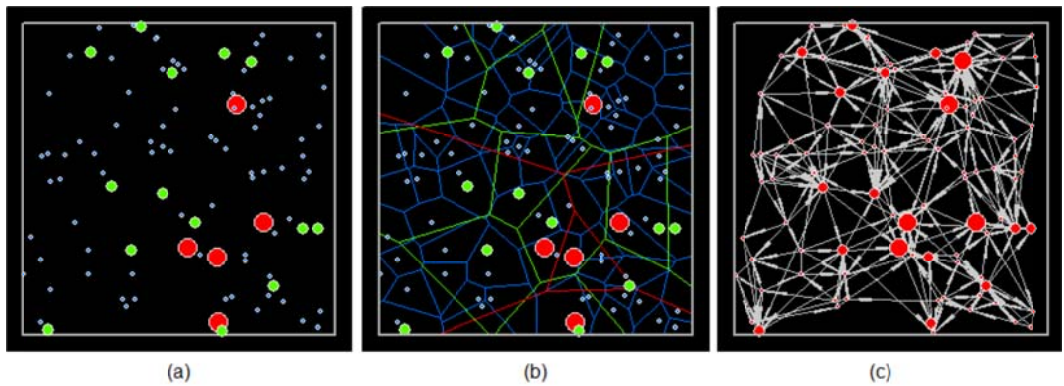

Figure 2: (Color online) The topological representation viewing cities as a whole
(Note: One hundred fictional cities with the sizes of 1, 1/2, 1/3, …, and 1/100 in three hierarchical levels, indicated by red, green and blue, are given some random locations (Panel a), the Thiessen polygons of the cities with the cities on the top (Panel b), and the complex network showing scaling hierarchy with dot sizes representing the degree of wholeness (Panel c). The cities in Panels a and b are differentiated from each other but lack local adaptation. The representation of cities in Panel c shows not only cities differentiation but also their adaptation to each other and to the whole network. In this connection, the point pattern in Panel c is more whole than that of those cities in Panels a and b.)

The topological representation is developed primarily for showing cities' structure from the perspective of topology and wholeness. However, it can be applied to carpets, city plans, building facades, and virtually any other designs. We can quantitatively examine if they are a living structure and further compute the degrees of beauty for individual coherent centers and the whole. When it is applied to designs, the key issue is to identify all latent centers, and their support relationships or topology, to set up a complex network representing the wholeness. Degrees of wholeness can then be computed relying on the mathematical model of wholeness (Jiang 2015b). The computed degrees of wholeness possess both properties of differentiation and adaptation, reflecting the underlying spatial configuration or arrangement of the centers. Alexander (2002–2005) referred to spatial configuration as geometry or living geometry. This should be more clearly called *topology*, hence the topological representation. The representation and topological analysis enable us to see the two transformation processes of differentiation and adaptation; see more details in the next section. The city sizes, or their equivalent occupied space of the Thiessen polygons, demonstrate differentiation, while the cities' growth and shrinkage show how cities adapt to their surroundings in the whole. We will further discuss these two transformation processes in Section 5. Before that, we apply the representation to some case studies.

**4. Living structures of China and UK natural cities**
We applied the topological representation to China (mainland) and UK (main island) natural cities to demonstrate that they are living structures. Natural cities in the context of this paper refer to naturally, objectively derived high-density patches using large amounts of streets nodes or tweeted locations, and based on head/tail breaks (Jiang 2015a). Natural cities, different from conventional concepts of cities defined or imposed by statistical or census authorities, provide a new instrument for better understanding urban structure and dynamics. In this study, we examine the cities' structure and, specifically, how cities of each country are differentiated from, and adapt to, each other to form a coherent whole from both statistical and topological perspectives. These perspectives enable us to see cities as a geometrically coherent whole, rather than an arbitrary assembly. We also developed some visualization methods to illustrate the underlying scaling hierarchy of the cities' structure. We adopted a recursive definition of geographic space, in which all cities of a country constitute a living structure, as do all hotspots of a city (Figure 3). We took China and UK each as a whole, as well as Beijing and London natural cities each as a whole, for detailed investigations.

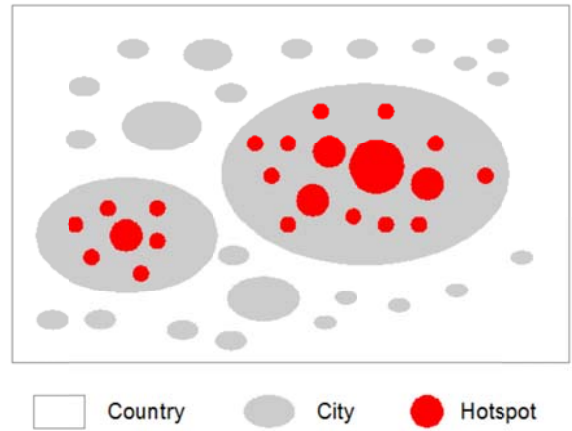

Figure 3: (Color online) A recursive definition of geographic space
(Note: Geographic space is not closed, which implies that a country is part of a continent or the globe.)

### 4.1 Data and data processing
The case studies are based on two datasets: 2,728,143 streets nodes in China (Long, Shen and Jin 2016), and 3,281,935 tweeted locations in the UK during June 1–8, 2014 (previously studied in Jiang et al. 2016) (Table 2). The 3 million locations were used to build up a huge triangulated irregular network (TIN) for each country. The TIN consists of far more short edges than long ones. The short and long edges respectively indicate high and low densities. All the short edges constitute individual patches called natural cities (see Jiang and Miao 2015 for a tutorial). After having extracted all the natural cities, we then took Beijing and London natural cities, respectively containing 157,755 and 203,986 locations, and each of the natural cities repeated the above processes to get individual hotspots in the cities. The hotspots to a city are what cities are to a country, so countries, cities, and hotspots constitute nested relationships (Figure 3).

Table 2: Numbers of locations and derived natural cities and hotspots

| Case studies | Locations | Cities/hotspots |
|---|---|---|
| China | 2,728,143 | 4322 |
| Beijing | 157,755 | 6054 |
| UK | 3,281,935 | 4664 |
| London | 203,986 | 8630 |

Following the procedure in Figure 2, these natural cities and hotspots were then topologically represented. The data processing was carried out mainly through ArcGIS and other related scripts for computing the degrees of wholeness. To make the paper self-contained, we briefly describe the data processing here. Use the head/tail breaks and base on city sizes to classify the centroids of the natural cities into different hierarchical levels. Create Thiessen polygons from the centroids, as shown in Figure 2. For convenience of ArcGIS users, this step can be done by: (1) ArcToolbox > Proximity > Create Thiessen Polygons (using a country's bounding box); and (2) Geoprocessing > Clip (using a country boundary). Write a simple script to create a complex network for all the Thiessen polygons; small cities point to adjacent large cities for those at a same level, and contained polygons point to containing polygons across two consecutive levels. Based on the complex network, compute the degrees of wholeness for individual Thiessen polygons or equivalently individual cities or hotspots.

Through the above data processing, we derived large amounts of natural cities: 67,700 for China and 124,608 for the UK. In order for the amounts of natural cities and hotspots to be of the same magnitude, we respectively chose the large cities by following the head/tail breaks for detailed investigation (Table 2) – the first head 4,322 large cities in China and the second head 4,664 large cities in the UK. Each of these cities or hotspots got two parameters: the size, measured by the number of locations; and the degree of wholeness.

### 4.2 Visualization of differentiation and adaptation

We found that cities are differentiated from each other in terms of the sizes and their degrees of wholeness. The differentiation can be characterized by a power law distribution and ht-index (Jiang and Yin 2014) for respectively measuring the hierarchy and the hierarchical levels. Figure 4 illustrates the natural cities in China, with the background and foreground respectively showing the six hierarchical levels of Thiessen polygons and the largest cities on the top three levels. We can clearly see the pattern of wholeness is just slightly different from that of sizes, indicating that sizes are well adapted to their surroundings. There are a few expanded cities, such as Shanghai and Shenyang, for they are surrounded or supported by many others. The pattern for the degrees of wholeness (Figure 4b) is considered the most harmonized and adapted. On the other hand, the pattern for city sizes (Figure 4a) resembles that of the degrees of wholeness. In fact, the city sizes and degrees of wholeness are well correlated with an R square of 0.82.

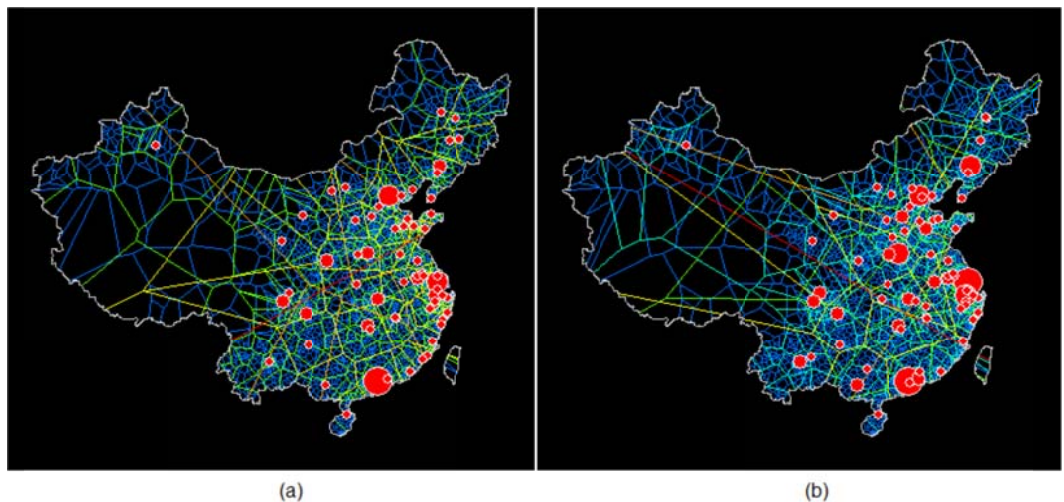

Figure 4: (Color online) China natural cities (4,322 in total) as a living structure from both cities' sizes and their degrees of wholeness
(Note: There are six hierarchical levels in terms of (a) the cities' sizes, with the 65 largest cities on the top four levels; and (b) the degrees of wholeness, with the 70 most beautiful cities on the top four levels.)

We visualized the complex network of cities with the foreground and background respectively representing links across scales or levels (blue lines) and within a same scale or level (gray lines) to further illustrate the underlying scaling hierarchy or the fact that cities are differentiated from and adapted to each other. The six hierarchical levels of China's natural cities are shown in Figure 5. Compared to the hierarchy based on the cities' sizes (Figure 5a), more cities are forced to the highest hierarchical levels in terms of the degrees of wholeness (Figure 5b). Therefore, there are three cities at the two highest levels in terms of the cities' sizes, whereas there are six at the two highest levels in terms of the degrees of wholeness.

We further examined the power distributions and ht-indices (Table 3). While examining p value in some cases in which the largest city or the hottest spot is too big (sometimes more than three times larger than the second largest city), we excluded it. This is to follow the concept of *primate city*, which was empirically well-observed and mathematically proven (Jefferson 1939, Chen 2012). Although there are two cases in which the p value is zero, their ht-indices are still very high around 5 and 6, indicating some heavy tailed distributions. It should be noted that the power law of cities' sizes is purely statistical, while the degrees of wholeness is not only statistical but also geometrical, indicating the property of adaptation. All these analytical results point to the fact that cities in a country constitute a living structure.

Table 3: Power law detection results and ht-indices for cities sizes and their degrees of wholeness
(Note: The power law detection is based on the robust method introduced in Clauset et al. (2009), while the ht-index is based on Jiang and Yin (2014))

| Case studies | Parameters | Alpha | Xmin | p | ht |
|---|---|---|---|---|---|
| China | sizes | 1.86 | 37 | 0 | 6 |
|  | wholeness | 2.14 | 0.0005103 | 0.18 | 6 |
| UK | sizes | 2.66 | 106 | 0.05 | 5 |
|  | wholeness | 2.28 | 0.0000487 | 0.07* | 6 |
| Beijing | sizes | 2.34 | 16 | 0.71 | 4 |
|  | wholeness | 2.44 | 0.0000430 | 0.05* | 6 |
| London | sizes | 2.20 | 23 | 0.04 | 6 |
|  | wholeness | 2.26 | 0.0000244 | 0 | 5 |

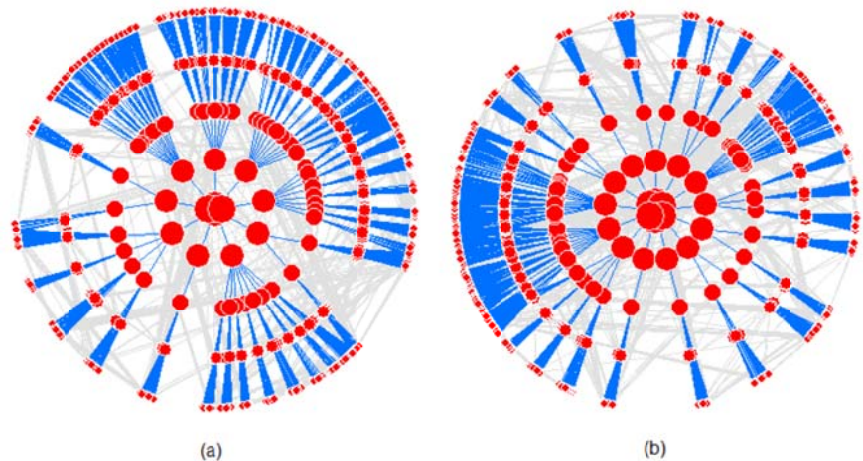

Figure 5: (Color online) Scaling hierarchy of China cities' sizes and their degrees of wholeness
(Note: Both the cities' sizes and their degrees of wholeness possess six hierarchical levels (a) for cities' sizes, and (b) for degrees of wholeness, indicating that the properties of differentiation and adaptation.)

The living structure appears not only at the country level, but also at the city level. To illustrate, the London natural city consists of 8,630 hotspots, the smallest of which are a patch with three locations. Both the sizes and the degrees of wholeness follow very well a power law distribution, with alpha being 2.20 and an ht-index greater than 5 (Table 3). The spatial distribution of some of the largest hotspots is illustrated in Figure 6a, with five at the two highest levels, and the most beautiful hotspots (Figure 6b) with eight at the highest level. Similar to Figure 5 at the country level, Figure 7 illustrates the complex network with the hierarchical tree in the foreground. It should be noted in Figure 6 that some small hotspots have the highest degrees of wholeness, thus becoming the most beautiful. This is understandable, since they get many supports from their surroundings, so become well-adapted to wholeness. The Beijing natural city has a similar structure, although the ht-index is only 4. Despite the fact that the two patterns in Figure 6 look dramatically different, the hotspot sizes correlate well with the degrees of wholeness. Except for the largest spot, the R square value is 0.87. To some extent, the R square indicates how cities' sizes adapted to each other; the higher the R square, the more adapted the cities.

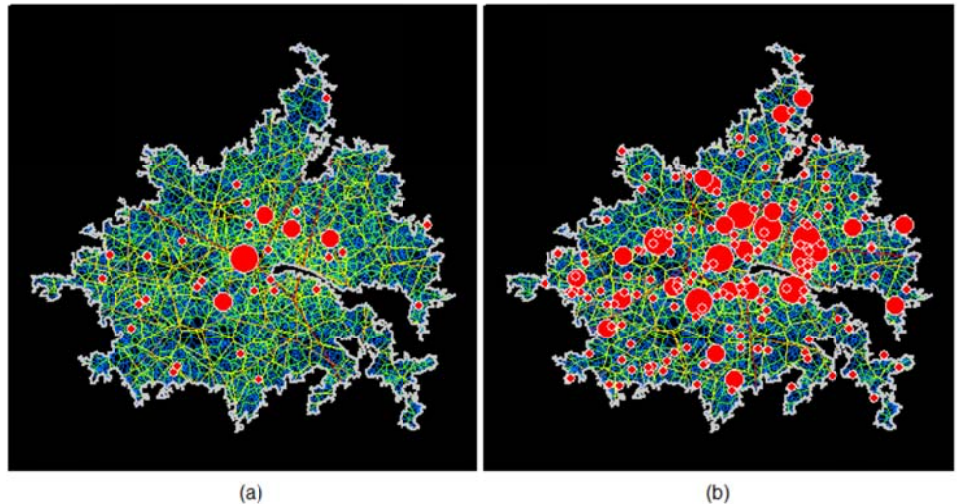

Figure 6: (Color online) London natural city consisting of 8,630 hotspots as a living structure from both cities' sizes and their degrees of wholeness
(Note: There are six hierarchical levels in terms of (a) the sizes of hotspots with the 32 largest on the top three levels, and five hierarchical levels according to (b) the degrees of wholeness with the 165 most alive spots on the top three levels.)

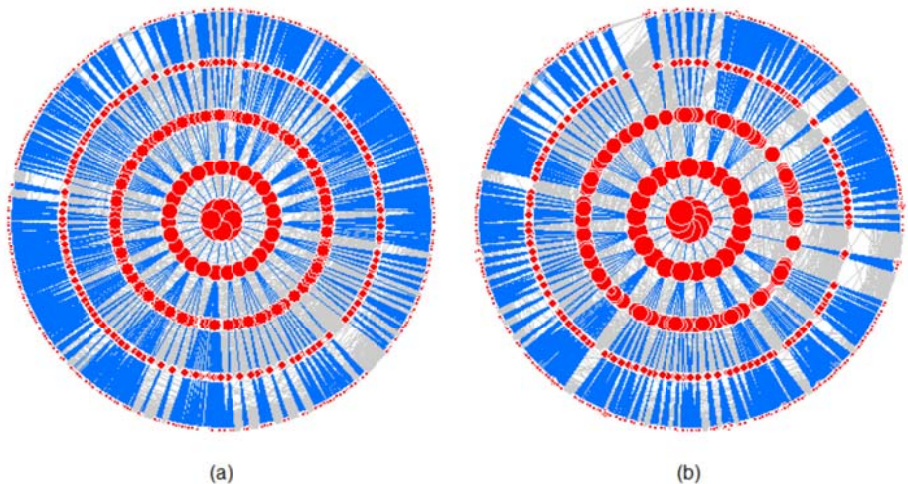

Figure 7: (Color online) Scaling hierarchy of London hotspot sizes and their degrees of wholeness
(Note: The hotspot sizes and their degrees of wholeness respectively possess (a) six and (b) five hierarchical levels, indicating the nature of differentiation and adaptation. Note that there are the top two levels at the center in (a).)

Through the case studies, we found that cities in a big country as a whole has a higher degree of wholeness than a city by itself. This is understandable, just as the human body is more complex than the human brain because the latter is part of the former. The results of the case studies provide further evidence to support the wholeness-oriented design, in particular the differentiation and adaptation processes. Through the topological representation and analysis, we illustrated and demonstrated that cities are a living structure, and so is a city by itself. The living structure illustrated at the city level is closely related to the concepts of vitality, imageability and legibility, which underlie the theory of the image of the city (Lynch 1960). These concepts, together with beauty, life, order and wholeness, all sound subjective, but are actually measurable and computable through the degrees of wholeness. For example, the image of the city can be formed simply because of the underlying living structure (Jiang 2013). There is little doubt that those largest or most beautiful nodes (representing cities or hotspots) in the inner circles of Figures 5 and 7 constitute the image of the country and city.

We can add a few remarks on the living structure to this point. First, the point pattern in Figure 2c is more whole than that in Figure 2a or 2b. This is because the degrees of wholeness under the influence of spatial configuration are more adapted to each other than to the cities' sizes. Second, with respect to Figures 5 and 7, the foregrounds are hierarchal trees, while the foregrounds and backgrounds together constitute complex networks (Jiang 2015c). In other words, cities are near decomposable in terms of Simon (1962), or cities are not trees but semi-lattices (Alexander 1965). Third, the results from the case studies point to the two properties of differentiation and adaptation at both country and city levels. According to Alexander (2002–2005), differentiation and adaptation, or the living structure in general, appears at all scales ranging from the Planck length ($10^{-43}$ m) to the scale of the universe itself ($10^{27}$ m), virtually from the infinitely small to the infinitely large. Based on these analytical insights, we further discuss and elaborate the implications of the topological representation and analysis.

## 5. Discussions on the topological representation and analysis

The topological representation is rooted in the living structure (Alexander 2002–2005), in the central place theory (Christaller 1933, 1966), and in fractal geometry (Mandelbrot 1982), as well as in the new definition of fractal: *a set or pattern is fractal if the scaling of far more small things than large ones recurs multiple times or with ht-index being at least three* (Jiang and Yin 2014). All these theories have one common thing in which they differ from Euclidean geometry, which essentially deals with regular or simple shapes. The geometry dealing with the living structure is referred to as *living geometry*, which *"follows the rules, constraints, and contingent conditions that are, inevitably, encountered in the real world"* (Alexander et al. 2012, p. 395). Beyond fractal geometry, living geometry aims not only for understanding things, but also for making things. Unfortunately, modernist urban planning and design is very much (mis-)guided by Euclidean geometric thinking (Mehaffy and Salingaros 2006). As a devastated result, modern architecture and city planning (Corbusier 1989) inevitably lead to dead or lifeless buildings and cities. Their design or structure lacks differentiation and adaptation, and is therefore not living structure. To create sustainable built environments, we therefore must abandon the Euclidean geometric thinking. What we want to abandon is the Euclidean geometric thinking rather than Euclidean geometry. Euclidean geometry is essential for fractal geometry, because one must first measure things in order to see far more small things than large ones. In this connection, Jiang and Brandt (2016) provide detailed arguments as to why the Euclidean geometric thinking is limited in understanding complex geographic forms and processes.

The analytical evidence and insights about the living structure developed in the present paper add further implications on the wholeness-oriented design. Any design or planning should respect the wholeness, and retain it and further extend it. This is the design philosophy Alexander (2002–2005) advocated through his theory of centers, in order to effectively create living structures in buildings, cities, and artifacts. This wholeness-oriented design philosophy applies not only to cities and buildings, but also to any artifacts in pursuing the living structure or beauty. Any design or structure must include many substructures across many scales that are differentiated from each other to form a scaling hierarchy. On the one hand, there is adaptation across all scales by a constant scaling ratio of approximately 2 or 3, also called scaling coherence (Salingaros 2005). On the other hand, the substructures must adapt to each other at same or similar scales. This design philosophy is in line with complexity theory, which states that cities emerged either autonomously or in some self-organized manners (e.g., Portugali 2000). The wholeness-oriented design reflects what humans, across different countries and cultures, did in the past centuries before the 1920s in creating buildings, cities, and artifacts. The two properties of differentiation and adaptation help us not only correctly conceive of, but also beautifully create architecture. If we differentiate a space in a way that treats it as a coherent whole, it would lead to a living structure for the space and even beyond towards a larger space. Practically speaking, wholeness-oriented design must be guided by the 15 properties discussed in Section 2. In this regard, many practical ways of creating living structure that are guided by scientific principles, rather than artistic standards, have been thoroughly discussed (e.g., Salingaros 2013, Mehaffy and Salingaros 2015).

The topological representation or the analysis in general helps us clearly understand some misperceptions in modern urban planning and design. First, order comes only from regular shapes or grids, governed by geometrical fundamentalism (Mehaffy and Salingaros 2006). In fact, what looks chaotic or disorderly on the surface could hold hidden or deep order. This hidden order is characterized by the scaling law of far more small things than large ones. This order recurs across all scales, ranging from the smallest to the largest. Second, order comes from a tree structure. This is a deadly misperception (Alexander 1965). As demonstrated in the above case studies and elsewhere (Jiang 2015c), a city is not a tree, but a complex network. In other words, a complex network is no less orderly than a tree. Different from a tree, parts in a complex network often overlap, nest, and interact each other. However, both complex networks and trees share the same scaling hierarchy of far more small things than large ones. The third misperception is that symmetry is always global. In fact, many living structures are full of local symmetries, yet lack global symmetry. For example, neither the Alhambra plan nor the Nolli map is globally symmetric. The global shape must adapt to both natural and built environments, and numerous local symmetries can enhance degrees of wholeness or order (Alexander 2002-2005). Fourth, cities are less orderly than the Sierpinski carpet; similarly, coastlines are less orderly than Koch curves. In fact, both cities and the Sierpinski carpet, or coastlines and Koch curves, share the same scaling or fractal or living or beautiful order: the former being statistical (Mandelbrot 1982), and the latter being rigid. To summarize and to paraphrase Alexander (2002-2005), the order in nature – natural systems, is essentially the same as that in what we build or make – buildings, cities or artifacts.

The two properties of differentiation and adaptation are governed by two fundamental laws: Scaling law and Tobler's law. Scaling law refers to some heavy tailed distributions for many natural and societal phenomena. It can be simply phrased as far more small things than large ones; or numerous smallest things, a few largest things, and some in between the smallest and the largest. The existing geometry-based representations cannot well address this scaling property of geographic features. Instead, the topological representation developed in this paper provides a useful instrument for seeing the scaling property. Tobler's law, or Tobler's first law of geography, refers to the effect of spatial autocorrelation or dependence, widely studied in the geography literature. Tobler's law states that *"everything is related to everything else, but near things are more related than distant things"* (Tobler 1970, p. 236). Current spatial analysis and spatial statistics in particular concentrates too much on the effect of spatial dependence, yet little on scaling law. Scaling law and Tobler's law complement each other for describing the Earth's surface: the former across all scales being global, while the latter on a single scale being local; the former on heterogeneity, while the latter on homogeneity; the former characterized by power law statistics and fractal geometry, as well as living geometry, while the latter by Gaussian statistics and Euclidean geometry. These two laws are fundamental not only to geographical phenomena, but also to any other living structure that recurs between the Planck length and the size of the universe itself.

**6. Conclusion**
Cities are not isolated, but are coherent entities within an interconnected whole. This situation is the same for a single city that consists of many coherent hotspots. This paper developed a topological representation that takes cities as a whole or hotspots as a whole to uncover their underlying scaling hierarchy, and to further understand the city as a problem in organized complexity (Jacobs 1961). We adopted a recursive definition of geographic space, under which a country acts as a coherent whole, as does any individual city. This paper provided analytical insights to advocate for wholeness-oriented design, particularly the differentiation and adaptation principles. To a great extent, this study provided scientific evidence for the claim that the order in nature is essentially the same as that in what we build or make (Alexander 2002–2005). The kind of topological representation and analysis can also be applied to modern architecture and urban design to objectively judge whether or not certain buildings and cities are stiff and lifeless in terms of the underlying structure. This is in line with what Alexander (2002-2005) claimed that the goodness of built environments is not a matter of opinion, but a matter of fact.

This paper clarified some misperceptions in urban design. The order or structure illustrated through the case studies cannot be characterized by Euclidean geometry or Gaussian statistics. For example, many living structures (except for carpets because of their rectangular shapes), lack global symmetry, but full of local symmetries. A complex network is no less orderly than a tree. To put it in another way, a city is not a tree, but a complex network (Alexander 1965, Jiang 2015c). Living structures are governed by two laws: Tobler's law at a same scale and scaling law across all scales, or equivalently the two spatial properties of dependence and heterogeneity. We further pointed out two types of coherence or adaptation: across all scales for those being far more small things than large ones, and at a same scale for those being more or less similar. The kind of topological representation can be applied to any works of art. The key issue is to identify the coherent entities or centers, and importantly their nested and overlapping relationships, so that a complex network can be set up for further computing their degrees of beauty or life. Ultimately, this paper provides analytical evidence and insights to support the wholeness-oriented design, particularly the two fundamental design processes of differentiation and adaptation.

**Acknowledgment**
I would like to thank Zheng Ren for his research assistance and Ding Ma for helping with some of the figures. The three anonymous referees and the editor deserve my special thanks for their constructive comments. However, any errors and inadequacies of the paper remain solely my responsibility.

**References:**
Alexander C. (1965), A city is not a tree, *Architectural Forum*, 122(1+2), 58-62.
Alexander C. (1993), *A Foreshadowing of 21st Century Art: The color and geometry of very early Turkish carpets*, Oxford University Press: New York.
Alexander C. (2002-2005), *The Nature of Order: An essay on the art of building and the nature of the universe*, Center for Environmental Structure: Berkeley, CA.
Alexander C., Neis H., Anninou A., and King I. (1987), *A New Theory of Urban Design*, Oxford University Press: London.
Batty M. and Longley P. (1994), *Fractal Cities: A geometry of form and function*, Academic Press: London.
Chen Y. (2012), The mathematical relationship between Zipf's law and the hierarchical scaling law, *Physica A*, 391(11), 3285-3299.
Christaller W. (1933, 1966), *Central Places in Southern Germany*, Prentice Hall: Englewood Cliffs, N. J.
Clauset A., Shalizi C. R., and Newman M. E. J. (2009), Power-law distributions in empirical data, *SIAM Review*, 51, 661-703.
Corbett J. P. (1979), *Topological Principles in Cartography*, Technical paper 48, U.S. Dept. of Commerce, Bureau of the Census: Washington D. C.
Corbusier L. (1989), *Towards a New Architecture*, Butterworth Architecture: London.
Egenhofer M. J. and Herring J. R. (1990), A mathematical framework for the definition of topological relationships, *Proceedings of the Fourth International Symposium on Spatial Data Handling*, International Geographical Union, Zurich 1990, 803–813.
Goodchild M. F. and Mark D. M. (1987), The fractal nature of geographic phenomena, *Annals of the Association of American Geographers*, 77(2), 265-278.
Goodchild M. F., Yuan M., and Cova T. J. (2007), Towards a general theory of geographic representation in GIS, *International Journal of Geographical Information Science*, 21(3), 239–260.
Jacobs J. (1961), *The Death and Life of Great American Cities*, Random House: New York.
Jefferson M. (1939), The law of the primate city, *Geographical Review*, 29(2), 226-232.

Jiang B. (2013), The image of the city out of the underlying scaling of city artifacts or locations, *Annals of the Association of American Geographers*, 103(6), 1552-1566.
Jiang B. (2015a), Head/tail breaks for visualization of city structure and dynamics, *Cities*, 43, 69-77. Reprinted in Capineri C., Haklay M., Huang H., Antoniou V., Kettunen J., Ostermann F., and Purves R. (editors, 2016), *European Handbook of Crowdsourced Geographic Information*, Ubiquity Press: London.
Jiang B. (2015b), Wholeness as a hierarchical graph to capture the nature of space, *International Journal of Geographical Information Science*, 29(9), 1632–1648.
Jiang B. (2015c), A city is a complex network, in: M. W. Mehaffy (editor, 2015), *Christopher Alexander A City is Not a Tree: 50th Anniversary Edition*, Sustasis Press: Portland, OR, 89-100.
Jiang B. (2016), A complex-network perspective on Alexander's wholeness, *Physica A*, 463, 475-484.
Jiang B. and Brandt A. (2016), A fractal perspective on scale in geography, *ISPRS International Journal of Geo-Information*, 5(6), 95; doi:10.3390/ijgi5060095
Jiang B. and Claramunt C. (2004), Topological analysis of urban street networks, *Environment and Planning B: Planning and Design*, 31(1), 151-162.
Jiang B. and Miao Y. (2015), The evolution of natural cities from the perspective of location-based social media, *The Professional Geographer*, 67(2), 295 - 306.
Jiang B. and Yin J. (2014), Ht-index for quantifying the fractal or scaling structure of geographic features, *Annals of the Association of American Geographers*, 104(3), 530–541.
Jiang B., Zhao S., and Yin J. (2008), Self-organized natural roads for predicting traffic flow: a sensitivity study, *Journal of Statistical Mechanics: Theory and Experiment*, July, P07008.
Köhler W. (1947), *Gestalt Psychology: An Introduction to New Concepts in Modern Psychology*, LIVERIGHT: New York.
Long Y., Shen Y. and Jin X. (2016), Mapping block-level urban areas for all Chinese cities, *Annals of the American Association of Geographers*, 106(1), 96–113.
Lynch K.(1960), *The Image of the City*, The MIT Press: Cambridge, Massachusetts.
Mandelbrot B. B. (1982), *The Fractal Geometry of Nature*, W. H. Freeman and Co.: New York.
McMaster R. B. and Usery E. L. (editors, 2005), *A Research Agenda for Geographic Information Science*, CRC Press: New York.
Mehaffy M. W. (editor, 2015), *Christopher Alexander A City is Not a Tree: 50th Anniversary Edition*, Sustasis Press: Portland, OR.
Mehaffy M. W. and Salingaros N. A. (2006), Geometrical fundamentalism, In: Salingaros N. A. (2006), *A Theory of Architecture*, Umbau-Verlag: Solingen.
Mehaffy M. W. and Salingaros N. A. (2015), *Design for a Living Planet: Settlement, science, and the human future*, Sustasis Press: Portland, Oregon.
Portugali J. (2000), *Self-Organization and the City*, Springer: Berlin.
Salingaros N. A. (2005), *Principles of Urban Structure*, Techne: Delft.
Salingaros N. A. (2013), *Algorithmic Sustainable Design: Twelve lectures on architecture*, Sustasis Press: Portland, Oregon.
Simon H. A. (1962), The architecture of complexity, *Proceedings of the American Philosophical Society*, 106, 468-482.
Theobald D. M. (2001), Topology revisited: representing spatial relations, *International Journal of Geographical Information Science*, 15(8), 689-705.
Tobler W. (1970), A computer movie simulating urban growth in the Detroit region, *Economic geography*, 46(2), 234-240.
Yuan M., Mark D. M., Egenhofer M. J. and Peuquet D. J. (2005), Extensions to Geographic Representations, in: McMaster R. B. and Usery E. L.(editors, 2005), *A Research Agenda for Geographic Information Science*, CRC Press: New York, 129 - 156.
Zipf G. K. (1949), *Human Behaviour and the Principles of Least Effort*, Addison Wesley: Cambridge, MA.